\documentclass[aps,prl,twocolumn,floatfix]{revtex4-1}
\usepackage{amsfonts,amsmath,amsthm,amssymb,amsopn,graphicx}
\usepackage{dsfont,bbm,xcolor}
\usepackage{enumitem}
\usepackage{hyperref}
\usepackage{breqn}
\usepackage{siunitx}
\usepackage{color}
\usepackage[]{breqn}

\makeatletter
\let\cat@comma@active\@empty
\makeatother
\newcommand{\AAA}{\mathcal{A}}
\newcommand{\BB}{\mathcal{B}}
\newcommand{\FF}{\mathcal{F}}

\newcommand{\ii}{ {\rm i} }
\def\bra#1{\mathinner{\langle{#1}|}}
\def\ket#1{\mathinner{|{#1}\rangle}}

\newcommand{\braket}[2]{\langle #1 \vert #2 \rangle}

\newcommand{\ave}[1]{{\langle #1\rangle}}

\newcommand{\tr}{\rm{tr}}

\def\one{\mathbbm{1}}
\def\bra#1{\mathinner{\langle{#1}|}}
\def\ket#1{\mathinner{|{#1}\rangle}}

\def\tr{{{\rm tr}}}

\def\one{\mathbbm{1}}

\usepackage{amsthm}
\theoremstyle{definition}

\begin{document}
\title{Out-of-Time-Ordered Crystals and Fragmentation}
\author{Berislav Bu\v{c}a}
\affiliation{Clarendon Laboratory, University of Oxford, Parks Road, Oxford OX1 3PU, United Kingdom}
\begin{abstract}
Is a spontaneous perpetual reversal of the arrow of time possible? The out-of-time-ordered correlator (OTOC) is a standard measure of irreversibility, quantum scrambling, and the arrow of time. The question may be thus formulated more precisely and conveniently: can spatially-ordered perpetual OTOC oscillations exist in many-body systems? Here we give a rigorous lower bound on the amplitude of OTOC oscillations in terms of a strictly local dynamical algebra allowing for identification of systems that are out-of-time-ordered (OTO) crystals. While OTOC oscillations are possible for few-body systems, due to the spatial order requirement OTO crystals cannot be achieved by effective single or few body dynamics, e.g. a pendulum or a condensate. Rather they signal perpetual motion of quantum scrambling. It is likewise shown that if a Hamiltonian satisfies this novel algebra, it has an exponentially large number of local invariant subspaces, i.e. Hilbert space fragmentation. Crucially, the algebra, and hence the OTO crystal, are stable to local unitary and dissipative perturbations. A Creutz ladder is shown to be an OTO crystal, which thus perpetually reverses its arrow of time. 

\end{abstract}		

\maketitle
{\it Introduction ---} A broken egg cannot unbreak itself, even though we may expend energy and produce heat to collect the pieces and mend it. This simple notion is an intuitive manifestation of the arrow of time. But can systems that are similar to eggs that spontaneously unbreak themselves exist? To answer this question we must turn to fundamental quantum mechanical consideration. More specifically, the question can be made more precise by studying the out-of-time-ordered correlation function (OTOC) that quantifies how much quantum information is lost (scrambled) in a many-body system. This quantity is a standard measure of the directionality of time \cite{OTOCArrowOfTime} and represents precisely the loss of information directly linked with R\'{e}nyi entropy growth \cite{OTOCentropy}. A system with persistently oscillating OTOC would thus arguably persistently and spontaneously reverse its own arrow of time. Such behaviour would go far beyond emergent persistent single (or few) body motion, e.g. a planet orbiting a star, or related limit cycle behaviour operating without time-periodic scrambling \cite{Strogatz_2000}.

All \emph{time crystals}, continuous \cite{Wilczek,Sacha2020,VedikaReview}, driven \cite{timecrystal1,timecrystal2,stochasticdiscrete,liang2020time}, dissipative \cite{Buca_2019,Booker_2020,dissipativeTCobs,Fabrizio1,Jamir1,Jamir2,timecrystalnew1,Seibold,Hadiseh}, and boundary \cite{Fazio,boundary2,boundary3,Lesanovsky,Shammah,Fabrizio2} are characterised by remarkable persistent oscillations of single or few body observables (and accompanying notions of continuous or discrete time translation symmetry breaking). By displaying non-trivial deterministic persistent dynamics, called \emph{non-stationary} dynamics \cite{Buca_2019}, time crystals seemingly violate basic ideas in quantum thermalization \cite{ETHReview}. Likewise, other systems, e.g. those with quantum scars \cite{scars,scars1,scars2,scars3,scarsRev,SanjayReview}, and others \cite{Dora,Benitez1,Benitez2,versteeg2020nonequilibrium}, also display non-stationary dynamics and have attracted lots of attention recently.

Motivated by the above considerations, in this Letter we define time-independent undriven systems that have stable non-trivially spatially dependent perpetual oscillations of the OTOC, with the OTOC becoming finite throughout the system, to be out-of-time ordered crystals (OTO crystals). Such spatially ordered dynamics of the OTOC can never be realised by any system with emerging effective single or few-body dynamics. We prove the existence of OTO crystals by providing a rigorous lower bound on the amplitude of oscillations of the OTOC in terms of a novel strictly dynamical symmetry algebra \cite{spinlace}. OTO crystals go far beyond standard time crystals because their oscillations signal a genuine reversal of scrambling (like an egg unbreaking itself) and entaglement spreading, rather than long-lived low friction motion of possibly effectively single body systems. This is to be contrasted with other notions of time crystals that are defined using local observables. OTO crystals do not violate the second law of thermodynamics. One may in fact think of an isolated many-body system acting as a bath for its own local degrees of freedom and imagine a possible fluctuation theorem for it \cite{FluctuationTheorem}, but any such theorem will not forbid either fluctuations or oscillations of entropy. In particular, OTOC oscillations imply oscillations in R\'{e}nyi entropy via \cite{OTOCentropy}. 

Remarkably, due to this novel strictly local dynamical algebra, such OTO crystals are completely stable to local unitary and even dissipative perturbations making the OTO crystal phase reasonably free from fine-tuning. This novel algebra goes beyond previously introduced dynamical symmetries \cite{Buca_2019} that in general do not form such an algebra. 
Moreover, this strictly local dynamical algebra is shown to directly imply that a Hamiltonian satisfying it has an exponentially large number of invariant local subspaces, i.e. Hilbert space fragmentation \cite{fragmentation1,fragmentation2,fragmentation3,fragmentationsubspace,feldmeier2021critically,Arnab}. In contrast to standard Hilbert space fragmentation, we call this \emph{local Hilbert space fragmentation} \cite{Dominik,fragmentationUN}, because the standard case is not seemingly due to such a structure. We will demonstrate all these principles by studying a spin-$1/2$ XYZ Creutz ladder that is shown to be an OTO crystal.  Finally, we argue that Hilbert space fragmentation by itself is not sufficient to guarantee perpetual oscillations of observables or OTOCs.

{\it Out-of-time ordered crystals ---}
We will focus on locally interacting lattice systems with Hamiltonians acting on some number of sites $N$ such that the full Hilbert space is $d^N$-dimensional with $d$ being the dimension of one site. These may be written as, $H=\sum_{|x-y|=\Lambda} h_{x,y},$ where the distance between the sites $x,y$ $\Lambda$ is finite. A strictly local operator is $B_{x}=b_{x,y}|_{|x-y|=\Lambda}$ because it acts non-trivially only a finite number of neighboring sites (between $x$ and $y$).

In order to proceed a useful concept will be dynamical symmetries \cite{Buca_2019}, which have been successfully applied for calculating dynamics of various non-stationary systems. Dynamical symmetries are \emph{extensive} operators $A$ satisfying $[H,A]=\omega A$. Systems for which this theory has been applied include isolated time crystals \cite{Marko1}, Floquet time crystals \cite{Marko2,Chinzei,Chinzei2}, dissipative time crystals \cite{Buca_2019,Booker_2020,DissipativeTCNew3}, synchronization \cite{Buca_2019,quantumsynch,buca2021algebraic}, and, when further extended, quantum many-body scars (e.g. \cite{scarsdynsym1,scarsdynsym2,scarsdynsym3,scarsdynsym4,scarsdynsym5,scarsdynsym6}). Dynamical symmetries, which should not be confused with those of the same name in classical physics \cite{classicaldynamicalsymmetries} and nuclear physics \cite{nucleardynamicalsymmetries}, are generalisations of spectrum generating algebras \cite{SGA} to the space of local or extensive operators. Instead of extensive, here we will assume strictly local dynamical symmetries (in the sense above).

Let us now precisely define an \emph{out-of-time ordered crystal} (OTO crystal) based on generalised out-of-time-ordered correlation functions,
\begin{equation}
C_{W_xV_y}(t_1,t_2)=-\ave{[W_x(t_1),V_y(0)][W_x(t_2),V_y(0)]}_\beta, \nonumber
\end{equation}
for some strictly local observables $W_x$ and $V_x$. Here the average is taken over the equilibrium state $\ave{\bullet}_\beta=\tr[\frac{e^{-\beta H}}{\tr[e^{-\beta H}]}\bullet]$. Intuitively, the OTOC quantifies information propagation from observable $W$ at point $x$ to $V$ at point $y$ at times $t_1,t_2$. Clearly, if the system is decoupled between site $x$ and $y$, then $C_{W_xV_y}(t_1,t_2)=0$. Generic behaviour for systems with finite local Hilbert spaces is saturation to some maximum \emph{constant} value \cite{Kukuljan} for finite $|x-y|$ in finite $t_1,t_2$. 

This motivates the definition of OTO crystals as those systems for which we have persistent oscillations at some frequencies $\omega_{1,2} \neq 0$ for some observables $W_x$ and $V_y$, i.e. the OTOC spectral function at frequencies $\omega_{1,2}$ is,
\begin{align}
&\FF_{W_xV_y}(\omega_1,\omega_2)= \label{oscicrit} \\
&\lim_{T \to \infty} |\frac{1}{T^2}\int^T_0 dt_1\int^T_0 dt_2 e^{\ii \omega_1 t_1+\ii \omega_2 t_2} C_{W_xV_y}(t_1,t_2)|>0, \nonumber
\end{align}
so that for \emph{all} $W_x,W_y$ with finite $|x-y|$ there are some finite times $t_1,t_2$ for which $C_{W_xV_y}(t_1,t_2)\neq 0$, and the number (or more generally measure) of $\omega_{1,2}$ for which $\FF_{W_xV_y}(\omega_1,\omega_2)$ is finite (non-zero) grows with distance $|x-y|$. The latter formal criterium means that the time-scales of the oscillations depend on the length scales, as expected from an interacting many-body effect because more frequencies enter at different distances. Moreover, we require that this property is stable to generic local perturbations, \emph{both} dissipative and unitary. 
 
The spatial modulation criterium of $\FF_{W_xV_y}(\omega_1,\omega_2)$ can never be met by a non-interacting system or a disconnected collection of finite systems because in those cases the OTOC would be identically $0$ for at least some $W_x,W_y$. The criteria, although at first glance rather formal, actually implies our desired requirement that oscillations in the OTOC indicate a perpetual increase and decrease of how much the quantum information is "scrambled" throughout the system \cite{OTOC1,scrambling1,scrambling2}. The stability criteria eliminates integrable systems, and ensures that the OTO crystal phase is not too fine-tuned. It should be however contrasted with discrete time crystals \cite{timecrystal2,timecrystal1,timecrystal3} that are stable to generic extensive perturbations. Our very strict criteria thus likely provide for the most similar behaviour to perpetual motion without violating the second law.

A sufficient criteria for persistent oscillations in the OTOC \eqref{oscicrit} is a simple closure condition given in \cite{SM}. To check the OTO crystal criterium it is sufficient to only give the lower bound in case $\omega_1=-\omega_2=\omega$, which we call $\FF_{WV}(x,y,\omega)$. In that case a novel OTOC spectral function lower bound is,
\begin{equation}
\FF_{WV}(x,y,\omega)\ge\ \AAA^\dagger \BB \AAA, \label{bound}
\end{equation}
where the vector,
\begin{align*}
&\AAA_k=-2 \sum_{m,\{z=1,2\}}(-1)^m\delta_{\omega,\omega_{km,z}} e^{-\beta( \delta_{z,2}\omega_k-\delta_{z,1}\omega_{km,z})}\times\\
&\left[\ave{W_x D_{k,m,z}}_\beta\right],
\end{align*}
with the crucial closure condition $\sum_m D_{k,m,1}=V_y A_k^\dagger$, $\sum_m D_{k,m,2}=A_k^\dagger V_y$, $[H,D_{k,m,z}]=\omega_{km,z}D_{k,m,z}$,  $\delta$ is the delta function and $k$ labels the distinct dynamical symmetries.This closure conditions states that overlap with dynamical symmetries is not enough for OTOC oscillations of observables $V_y$. $V_y$ must close under multiplication with a strictly dynamical symmetry $ A_k$ into another set of dynamical symmetries $D_{k,m,z}$. The matrix $\BB_{k,j}=\ave{A^\dagger_k A_j}_\beta$ is Hermitian. The expression is seemingly complicated, but is intuitive. It says that the bound is 0 if $W_x$ has the same overlap with $V_yA^\dagger_k$ and $A_k^\dagger V_y$. This immediately implies that it is 0 for non-interacting systems, as needed, and implies that $A_k$ needs to be an operator that is genuinely many-body, i.e. connecting $x$ and $y$ beyond the disconnected correlator level. This is an algebraic manifestation of the spatial modulation requirement above. The expression simplifies considerably at infinite temperature $\beta=0$. 

By construction, strictly local dynamical symmetries commute with any operator that is outside their support, i.e. $[A_k,X]=0$, if $X$ acts trivially on the support of $A_k$ $\Lambda$. Therefore, their existence is stable to any such local perturbation whether dissipative or unitary. However, remarkably, further stability is also possible, as we will see later in the example.

{\it Fragmentation, frustration and strictly local dynamical symmetries ---} 
Before moving on to the example, let us discuss the connection to fragmentation. Hilbert space fragmentation \cite{fragmentation3} is characterised by decomposition of the Hamiltonian into an exponential number of spatially disconnected invariant subspaces. The dimensions of these subspaces may range from finite to those whose size grows exponentially with the system size. 
Consider the influence of this on the time evolution of a local observable $\ave{O(x,t)}$ in a large, but finite system,
\begin{equation}
\ave{O(x,t)}=\sum_{j,k} e^{\ii t (E_j-E_k) }\bra{E_j}O_x\ket{E_k}\braket{\psi}{E_j}\braket{\psi}{E_k}^{*},  \nonumber
\end{equation}
where $H\ket{E_k}=E_k \ket{E_k}$ and we have assumed that the Hamiltonian is block diagonalized. If the blocks do not have special structure, an initial state will connect all blocks for generic observables. Even if the initial state is contained in only one exponentially large block, one may still invoke eigenstate dephasing \cite{eigenstatedephasing} within the block itself that will lead to relaxation. If the block is finite, then dephasing is not possible inside the block. However, eigenstate thermalization  which states that off-diagonal terms $\bra{E_j}O_x\ket{E_k}, j\neq k$, will be exponentially surpressed in system size, implies generically thermalization and relaxation even for finite blocks. 

If, however, the subspaces are \emph{local}, i.e. there is a projector $P^{n}_x$ to a necessarily finite eigenspace $\ket{E_n}$, which is strictly local, then clearly this immediately implies the existence of a strictly local conservation law $[H,P^{n}_x]=0$, rendering the system non-ergodic. This we call \emph{local fragmentation}.

Importantly, this alone is \emph{not} sufficient to guarantee persistent oscillations because the eigenspace may be degenerate (e.g. due to flat bands \cite{strictlylocalsymmetries,Kuno,Kuno2,flatband}). If the eigenspace is not degenerate, then operators of the form $A_x=\one \otimes \ket{E_j}\bra{E_k} \otimes \one$ are strictly local dynamical symmetries satisfying, $[H,A_x]=(E_j-E_k) A_x,$ \cite{spinlace}. The $\one$ denotes the rest of the system on which the operator acts trivially, which, for future purposes of studying the Creutz ladder, we split into a left and right part. By previous arguments \cite{Marko1,Marko2}, the existence of such operators implies persistent oscillations of observables. Conversely, if there exists a strictly local dynamical symmetry $A_x$, there is a strictly local conservation law $Q_x=[A_x,A_x^\dagger]$, that may be diagonalized. If $Q_x$ exists for a number of $x$ that is a finite fraction of the volume, then fragmentation into local subspaces happens. However, in true fragmentation such subspaces are \emph{not} in general local. Therefore, it is useful to check whether a fragmented models has dynamical symmetries, as exact results are then possible for quenches \cite{Marko1}, correlation functions \cite{Marko1,Marko2} and, as we shall see here, OTOCs. Finally, similarity between fragmentation and strictly local weak symmetries \cite{BucaProsen} has also been noted for open quantum systems \cite{Essler}. 

Local fragmentation has been associated with frustration in \cite{Dominik}. However, frustration is not needed for local fragmentation. In fact, what is sufficient is the existence of a strictly local conservation law $R_x$ (not to be confused from the one we called $Q_x$) that has a \emph{non-degenerate subspace}. From now on we will include all conservation laws $Q_x$ into the set of dynamical symmetries $A_x$ because they are special cases of dynamical symmetries with frequency 0. 

\begin{figure}[b]
\begin{center}
\includegraphics[scale=0.45]{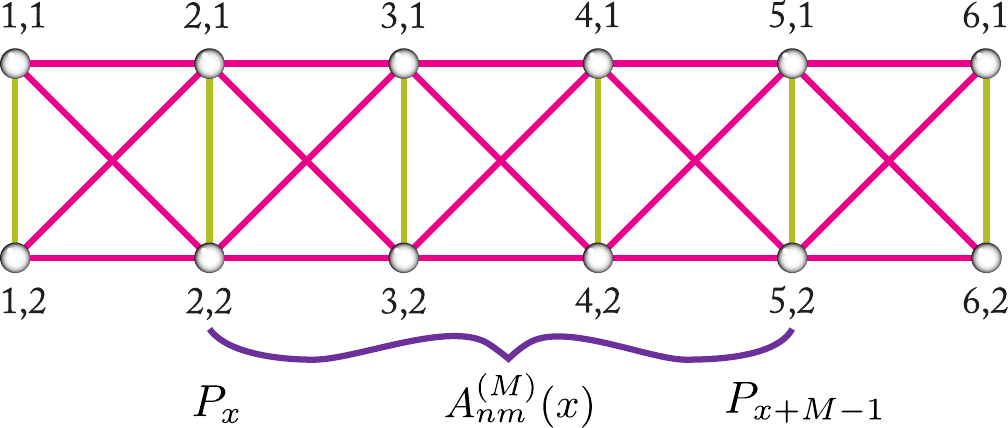}
\end{center}
\vspace{-4.3mm}
\caption{The spin-$1/2$ XYZ Cruetz ladder with $\Delta_\alpha$ (yellow) and $\Delta^{'}_\alpha$ (pink) interactions. The model is reflection symmetric along each rung $x$. Strictly local dynamical symmetries $A^{(M)}_{nm}(x)$ contain projectors $P_x/P_{x+M-1}$ on their boundaries defined in the text.}
\label{fig1}
\end{figure}

{\it Example: spin-$1/2$ XYZ Creutz ladder ---}
The spin-$1/2$ XYZ Creutz ladder exhibits local fragmentation. It is given by the Hamiltonian,
\begin{equation}
H=J\sum_{x,\alpha} \Delta_\alpha s^\alpha_{x,1}s^\alpha_{x,2}+\Delta^{'}_\alpha (s^\alpha_{x,1}+s^\alpha_{x,2})(s^\alpha_{x+1,1}+s^\alpha_{x+1,2}),
\end{equation}
where $s^\alpha_{x,y}$, $\alpha=x,y,z$ is a spin-$1/2$ Pauli matrix acting on the site indexed by the rung $x=1\ldots N$ and leg $y=1,2$ (see Fig.~\ref{fig1}). 

The model is not frustrated, but has a strictly local reflection symmetry $R_x$ on each rung $x$. Crucially, the $-1$ subspace of this operator is not degenerate, with projector $P_x=\one_{2^{x-1}}\otimes (\ket{\uparrow\downarrow}-\ket{\downarrow\uparrow})(\bra{\uparrow\downarrow}-\bra{\downarrow\uparrow})\otimes\one_{2^{N-x}}$ and $\one_{n}$ being the $n \times n$ identity matrix. Take any $2M-$site subsystem $\Lambda_{M,x}$ ($M$ neighboring rungs to the right of rung $x$) with Hamiltonian $H^{(M)}_x$. It is easy to see due to the aforementioned conservation of $P_x$ and the trivial topology of the ladder, any operator of the form $A^{(M)}_{nm}(x)=P_{x}\ket{n}_{M-2}\bra{m}_{M-2} P_{x+M-1}$ with $H^{(M)}) \ket{k}_M=E_{k} \ket{k}_M$ is a strictly local dynamical symmetry with $[H,A^{(M)}_{nm}(x)]=(E^{(M)}_n-E^{(M)}_m)A^{(M)}_{nm}(x)$. Note the existence of a strictly local symmetry $R_x$ with non-degenerate subspace is neither necessary \cite{inprep} nor sufficient \cite{spinlace} to obtain strictly local dynamical symmetries. 

Importantly, the existence of dynamical symmetries immediately implies that observables do not relax, even at infinite temperatures and for quenches from generic initial states. This means that the system has a stronger property than quantum many-body scarring (cf. \cite{fragscars}). It also implies the existence of l-bits (or strictly local conservation laws) $[A^{(M)}_{nm}(x),A^{(M')}_{m'n'}(y)]$ for $\omega^{(M)}_{mn}=\omega^{(M')}_{m'n'}$ with $\omega^{(M)}_{mn}=E^{(M)}_m-E^{(M)}_n$. These generalise the ones studied in \cite{flatband} and split the system along $-1$ sectors. 

\begin{figure}[!htb]
\begin{center}
\includegraphics[scale=0.65]{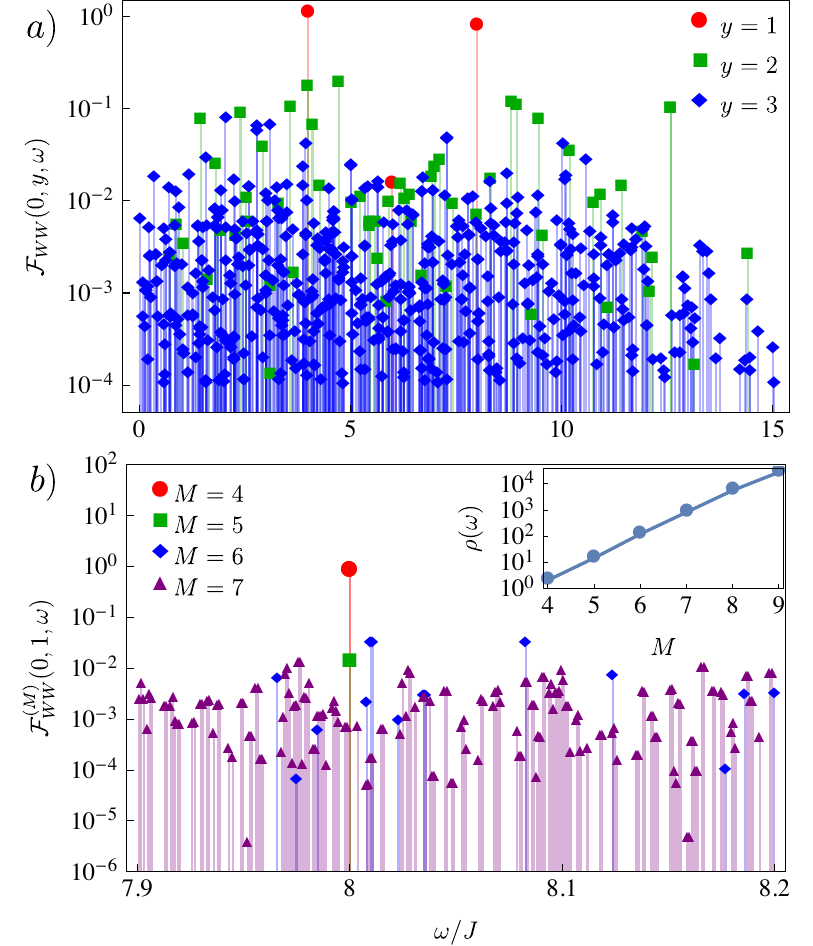}
\end{center}
\vspace{-4.4mm}
\caption{a) The lower bound for the OTOC spectral function for $W_{x/y}=s^z_{x/y,1}s^z_{x/y,2}$ for various separations $y$ using $A^{(M)}_{nj}(z)$ with the smallest non-trivial $M$. b) The same bound for $y=1$ and various $M$. The inset confirms that the density of \emph{distinct} frequencies scales exponentially with $M$ implying that larger $M$ become quickly irrelevant due to dephasing. Data for $\Delta_x=\Delta_y=2$, $\Delta_z=\Delta^{'}_x=1$, $\Delta^{'}_y=\Delta^{'}_z=1/2$.}
\label{fig2}
\end{figure}

We may also easily apply the bound \eqref{bound} to an observable $V_y$ by picking a subsystem of $M$ sites and writing $V_y$ in the eigenbasis of $H^{(M)}_x$. Focusing only on one choice of $H^{(M)}_z$ acting non-trivially only on $\Lambda_{M,z}$ at infinite temperature $\beta=0$ the lower bound $\FF^{(M,z)}_{WV}(x,y,\omega)$ takes the compact form (using orthonormality of the eigenstates of $H^{(M)}_z$),
\begin{align}\label{boundC}
&\FF^{(M,z)}_{WV}(x,y,\omega)=\\
&4\tilde{\sum}_{m,n}|\tilde{\sum}_{k,j}v_{kj}\tr[W_x ( A^{(M)}_{nj}(z)-A^{(M)}_{km}(z))]|^2,\nonumber
\end{align}
with $v_{kj}=\tr(V_y A_{jk})$ being chosen to have the same phase, and where the sums $\tilde{\sum}$ go over only those $n,j$ and $j,m$ such that $\omega=\omega_{M,nj}=\omega_{M,km}$, $\omega_{M,nm}:=E^{(M)}_n-E^{(M)}_m$. By decomposing the various subspaces of $H$ into the $\{+1,-1\}^{\otimes N}$ subspaces of all the $R_x$ we can ensure orthonormality of all the overlapping $A^{(M)}_{nj}(z)$ strictly local dynamical symmetries. The maximal bound is therefore the simple sum over the bounds of each subspace from \eqref{boundC}. 
We plot the OTOC spectral function bound for a many-body observable $W_x=V_x=s^z_{x,1}s^z_{x,2}$, for several separations $y-x$ (Fig.~\ref{fig2}) and for the minimal $M$ that has finite overlap with $W_x,V_y$. 
The larger the separation the more frequencies come into the bound showing that it is indeed a OTO crystal according to our definition. As numerically observed in \cite{Dominik}, despite localization, the OTOC becomes finite for finite $|x-y|$. Note the difference to MBL that does not have persistent OTOC oscillations because of dephasing of l-bits \cite{Fan_2017}. To see how fast the bound saturates in Fig.~\ref{fig2}~$b)$ we compare the spectral OTOC bound coming from various $M$ for $y-x=1$. As $M$ is increased, the corresponding dynamical symmetries $A^{(M)}_{nm}(x)$ become less local and have lower overlap with local operators.

More frequencies come into the dynamics, which become exponentially more dense and incommensurate (as seen in the inset of Fig.~\ref{fig2}~$b$), as expected due to the exponential growth of dimension of $H^{(M)}$. This leads to eigenstate dephasing \cite{eigenstatedephasing}, i.e. destructive interference, indicating that only a few $M$ are sufficient to saturate the bound. This also explains the numerical observation of \cite{Dominik} that the smallest subspaces contribute signficantly to the OTOC in particular in the long-time limit: the bound presented here is by construction an infinite time average. The model is even more stable than OTO crystals generally because the only local perturbations that can break $A^{(M)}_{nm}$ are those that break the local reflection symmetry $R_x$. The OTO crystal is thus further protected by this symmetry. 

{\it Conclusion ---} 
We have shown that quantum many-body systems can spontaneously reverse their own arrow of time, as quantified by OTOC \cite{OTOCArrowOfTime,OTOCentropy}, without violating the laws of thermodynamics. We call such systems OTO crystals - their out-of-time-ordered correlations display both finite frequency oscillations and non-trivial spatial dependence of these frequencies. OTO crystals are stable to both local unitary and dissipative perturbations. OTO crystals cannot be formed by non-interacting models or with one single effective degree of freedom. Rather than having such trivial dynamics at constant entropy, they have oscillations of many-body scrambling itself. We have rigorously shown that the OTO crystal behaviour is implied by the existence of strictly local dynamical symmetries that were used to bound the amplitude of the OTOC oscillations. Furthermore, strictly local dynamical symmetries imply a type of \emph{local} Hilbert space fragmentation, that should be contrasted with true fragmentation \cite{fragmentation3}, which does not seemingly guarantee persistent oscillations by itself. However, whether such oscillations are possible without strictly local dynamical symmetries remains an open question in particular in light of the OTOC study in \cite{feldmeier2021critically}. Connecting true fragmentation with the existence of conservation laws \cite{SanjayNew} could be used to study such a question. A simple example of an OTO crystal, a spin-$1/2$ XYZ Creutz ladder, was given. It essentially partitions itself into blocks of increasing size that partially constrains quantum information flow causing it to move back and forth, until dephasing is achieved for large enough blocks.

The results open many other interesting avenues for further exploration. Importantly, can recent highly successful hydrodynamic theories \cite{GHD1,GHD2,ampelogiannis2021ergodicity}, account for OTO crystals? The introduced notion of stability to local perturbations is reminiscent of topological quantum order \cite{topologicalquantumorder}, but, it is now present even at finite temperatures and frequencies. Can this similarity be exploited to have dynamical topological quantum information processing at finite temperatures in interacting systems? Likewise, oscillating entanglement could have important implications for quantum information processing and metrology \cite{metrology}. Another general question is what do oscillations in the OTOC imply for entanglement \cite{Olalla} and transport \cite{DissipativeTCNew3}.  Finally, OTO crystals could be identified in other systems, even though all the known non-stationary quantum many-body systems (e.g. \cite{Buca_2019,spinlace,Marko1}) posses only one frequency of oscillations and do not display non-trivial spatial dependence in that frequency. Can OTO crystals be achieved without constrained dynamics? In particular, it would be interesting to explore gravitational waves in holographically dual models displaying such oscillations \cite{Saso,PhysRevX.11.031003}. Connections to \emph{spontaneous} dynamical phase transitions \cite{dynphas} will be explored. Finally, a possibly experimental implementation using a bosonic Creutz ladder with circuit QED could simulate our spin-$1/2$ model in the hard-core limit of the bosonic system \cite{HadisehCreutz}. Likewise, experimental protocols for measuring OTOC exist, e.g. \cite{ReyOTOC,OTOCArrowOfTime}, making our OTO crystal potentially experimentally relevant
\begin{acknowledgments}
{\it Acknowledgements ---} Assistance from V. Juki\'{c} Bu\v{c}a with Figure 1 and useful comments from C. Booker, O. Castro Alvaredo, J. Feldmeier, D. Jaksch, S. Moudgalya, T. Prosen, S. Sarkar and A. Sen are very much acknowledged. Funding from EPSRC programme grant EP/P009565/1, EPSRC National Quantum Technology Hub in Networked Quantum Information Technology (EP/M013243/1), and the European Research Council under the European Union's Seventh Framework Programme (FP7/2007-2013)/ERC Grant Agreement no. 319286, Q-MAC is gratefully acknowledged.
\end{acknowledgments}

\bibliographystyle{apsrev4-1}
\bibliography{main}
\widetext
\clearpage
\begin{center}
\textbf{\large Supplementary Material:  Out-of-Time-Ordered Crystals and Fragmentation}\\
\end{center}
\setcounter{equation}{0}
\setcounter{figure}{0}
\setcounter{table}{0}
\setcounter{page}{1}
\makeatletter
\renewcommand{\theequation}{S\arabic{equation}}
\renewcommand{\thefigure}{S\arabic{figure}}

In this Supplemental Material the proof of the OTOC spectral function bound from the main text is given.

\section*{A--- Proof of the OTOC spectral function bound}
The proof proceeds similarly to \cite{Marko2}. We begin by defining,
\begin{equation}
\mathcal{G}=\frac{1}{T}\int_0^T dt e^{\ii \omega t}[ W(t),V]+\sum_k \alpha_k A_k,
\end{equation}
where $A_k$ are a set of dynamical symmetries, $\alpha_k$ are complex coefficients and we suppressed the spatial dependence of $W,V$ for neatness of notation. We have (likewise suppressing $\beta$),
\begin{equation}
\ave{\mathcal{G}^\dagger \mathcal{G}} \ge 0 \label{basicbound},
\end{equation}
with the average being taken over an equilibrium state $\ave{\bullet}=\tr[\frac{e^{-\beta H}}{\tr[e^{-\beta H}]}\bullet]$. This gives,
\begin{align}
&\ave{\mathcal{G}^\dagger \mathcal{G}}=\label{S2} \\\nonumber
&\frac{1}{T^2} \int^T_0 dt_1 \int^T_0 dt_2 \left\{e^{-\ii \omega(t_1-t_2)}\ave{[W(t_1),V][W(t_2),V]}- \sum_k \left[e^{-\ii \omega t_1}\alpha_k^{*}\ave{A^\dagger_k [W(t_1),V]}+c.c.\right]+\sum_{k,m} \alpha_k^{*}\alpha_m \ave{A^\dagger_k A_m} \right\}.
\end{align}
In order to eliminate the time-dependence of $W(t_1)$ in the second term in \eqref{S2} we, as in the main text, assume that we can express $\sum_m D_{k,m,1}=V_y A_k^\dagger$, $\sum_m D_{k,m,2}=A_k^\dagger V_y$, $[H,D_{k,m,z}]=\omega_{km,z}D_{k,m,z}$. Furthermore, we use cyclicity of the trace, time invariance of the $\rho$, and the exponential form of the dynamical symmetry condition $e^{\ii H t} A_k e^{-\ii H t}=e^{\ii \omega_k t} A_k$ to get,
\begin{equation} \label{S4}
\lim_{T \to \infty}\frac{1}{T}\int_0^T dt_1 e^{\ii \omega t_1}
\alpha_k^{*}\ave{A^\dagger_k [W(t_1),V]}=\alpha_k^{*}\sum_m\left[e^{\omega_{km,1}(\beta)}\delta_{\omega,\omega_{km,1}}\ave{W D_{k,m,1}}-e^{-\omega_{km,2}(\beta)}\delta_{\omega,\omega_{km,2}}\ave{W D_{k,m,2}}\right],
\end{equation}
where the $\delta$ function is obtained by time-averaging,
\begin{equation}
\delta_{\omega,\omega_{km,z}}=\lim_{T \to \infty}\frac{1}{T}\int_0^T dt e^{\ii (\omega-\omega_{km,z})t}.
\end{equation}
We plug \eqref{S4} back into \eqref{S2}. Then, using \eqref{basicbound} we can find a lower bound on,
\begin{equation}\label{spectdef}
\mathcal{F}=\frac{1}{T^2} \int^T_0 dt_1 \int^T_0 dt_2 e^{-\ii\omega(t_1-t_2)}\ave{[W(t_1),V][W(t_2),V]},
\end{equation} 
by finding the extremum of the rest of \eqref{S2}. To find this, like in \cite{Marko2}, we take derivative over $\alpha_k^{*}$ and equate to 0,
\begin{equation*}
\frac{d}{d \alpha_k^{*}}\left[\ave{\mathcal{G}^\dagger \mathcal{G}}-\mathcal{F}\right]=0,
\end{equation*}
which is easily solved and gives,
\begin{equation} \label{altbound}
\frac{1}{T^2} \int^T_0 dt_1 \int^T_0 dt_2 e^{-\ii\omega(t_1-t_2)}\ave{[W(t_1),V][W(t_2),V]} \ge \AAA^\dagger \BB \AAA,
\end{equation} 
where, as in the main text,
\begin{equation}
\AAA_k=-2 \sum_{m,\{z=1,2\}}(-1)^m\delta_{\omega,\omega_{km,z}} e^{-\beta( \delta_{z,2}\omega_k-\delta_{z,1}\omega_{km,z})}\left[\ave{W_x D_{k,m,z}}\right],
\end{equation}
and the Hermitian matrix,
\begin{equation*}
\BB_{k,j}=\ave{A^\dagger_k A_j},
\end{equation*}
which gives the bound from the main text. 

\bibliographystyle{apsrev4-1}
\bibliography{OTO_crystal_SM}

\end{document}